\documentclass[runningheads]{llncs}

\usepackage[table]{xcolor}
\usepackage{tikz}
\pgfdeclarelayer{back}
\pgfsetlayers{back,main}

\usepackage{eqparbox}
\newbox\nodebox
\usetikzlibrary{trees,shapes.geometric,calc,fit,petri,positioning}
\usepackage{makecell}

\usepackage{graphicx}
\usepackage{hyperref}
\usepackage{url}

\usepackage{amssymb}
\usepackage{csquotes}
\usepackage[official]{eurosym}
\usepackage{multirow}
\usepackage{subcaption}
\usepackage{todonotes}

\usepackage{tikz}
\usetikzlibrary{fit}

\begin{document}
%
\title{Cortado---An Interactive Tool for Data-Driven Process Discovery and Modeling}
\titlerunning{Cortado --- An Interactive Process Discovery Tool}
%
\author{Daniel Schuster\inst{1}\orcidID{0000-0002-6512-9580} \and
Sebastiaan J. van Zelst\inst{1,2}\orcidID{0000-0003-0415-1036} \and
Wil M. P. van der Aalst\inst{1,2}\orcidID{0000-0002-0955-6940}}

\authorrunning{D. Schuster et al.}
%
\institute{Fraunhofer Institute for Applied Information Technology FIT, Germany\\
\email{\{daniel.schuster,sebastiaan.van.zelst\}@fit.fraunhofer.de}
\and
RWTH Aachen University, Germany\\
\email{wvdaalst@pads.rwth-aachen.de}}
\maketitle              
\begin{abstract}
Process mining aims to diagnose and improve operational processes. 
Process mining techniques allow analyzing the event data generated and recorded during the execution of (business) processes to gain valuable insights.
Process discovery is a key discipline in process mining that comprises the discovery of process models on the basis of the recorded event data.
Most process discovery algorithms work in a fully automated fashion.
Apart from adjusting their configuration parameters, conventional process discovery algorithms offer limited to no user interaction, i.e., we either edit the discovered process model by hand or change the algorithm's input by, for instance, filtering the event data.
However, recent work indicates that the integration of domain knowledge in $\textnormal{(semi-)automated}$ process discovery algorithms often enhances the quality of the process models discovered.
Therefore, this paper introduces Cortado, a novel process discovery tool that leverages domain knowledge while incrementally discovering a process model from given event data.
Starting from an initial process model, Cortado enables the user to incrementally add new process behavior to the process model under construction in a visual and intuitive manner.
As such, Cortado unifies the world of manual process modeling with that of automated process discovery.


\end{abstract}
\section{Introduction}
Process mining techniques allow analyzing the execution of (business) processes on the basis of event data collected by any type of information system, e.g., SAP, Oracle, and Salesforce. 
Next to \emph{conformance checking} and \emph{process enhancement}, \emph{process discovery} is one of the three main sub-disciplines in process mining~\cite{DBLP:books/sp/Aalst16}.
Process discovery aims to learn a process model from observed process behavior, i.e., \emph{event data}.
Most process discovery algorithms are fully automated.
Apart from adjusting configuration parameters of a discovery algorithm, which, for instance, can influence the complexity and quality of the resulting models, the user has no direct option to steer or interact with the algorithm.
Further (indirect) user interaction is limited to either changing the input, i.e., the event data fed into the discovery algorithm, or manipulating the output, i.e., the discovered process model.
Thus, conventional process discovery algorithms work like a \emph{black box} from the user's perspective.

Several studies indicate that exploiting domain knowledge within \mbox{(semi-)} automated process discovery leads to better process models~\cite{DBLP:conf/er/DixitVBA18,DBLP:conf/bpm/BeneventoDSAA19}.
Recent work has proposed the tool ProDiGy~\cite{DBLP:conf/rcis/DixitBA18}, allowing the user to interact with automated process discovery.
However, the tool approaches the user-interaction from a \emph{modeling-perspective}, i.e., a human modeler supported by the underlying algorithms (including an \emph{auto-complete} option) is central to the tool and makes the design decisions for the model.
Thus, model creation is still a largely manual endeavor.

\begin{figure}[tb]
    \centering
    \includegraphics[width=\textwidth, trim=0cm 8.2cm 0cm 0cm,clip]{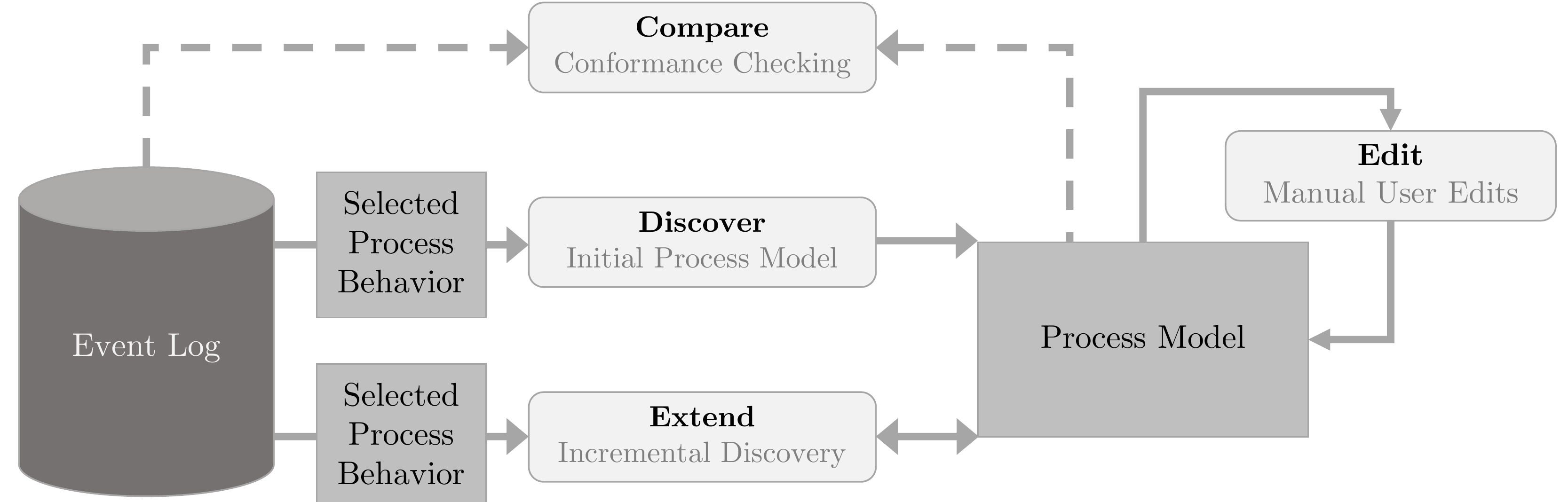}
    \caption{Overview of Cortado's core functionality. The user discovers an initial model from user-selected process behavior. Next, the obtained process model can be incrementally extended by new process behavior from the event log. In addition, the user can edit the process model anytime and compare it with the event log}
    \label{fig:overview}
\end{figure}

This paper introduces Cortado, an interactive tool for data-driven process discovery and modeling.
Cortado exploits automated process discovery to construct process models from event data in an incremental fashion.
Main functionalities of our tool are visualized in \autoref{fig:overview}.
The central idea of Cortado is the incremental discovery of a process model, which is considered to be \enquote{under construction}.
Cortado thereby utilizes the user's domain knowledge by delegating the decision to the user, which is about selecting the observed process behavior that gets added to the process model.

Cortado allows for discovering an initial process model from a user-selected subset of observed process behavior with a conventional process discovery algorithm (see \textbf{Discover} in \autoref{fig:overview}).
Alternatively, one can also import a process model into Cortado.
Cortado allows incrementally extending an initially given process model, which is either imported or discovered, by adding process behavior that is not yet described by the process model \enquote{under construction}.
Thus, the user is required to incrementally select process behavior from the event log and to perform incremental process discovery.
Our incremental discovery algorithm~\cite{10.1007/978-3-030-50316-1_25} takes the current process model and the selected process behavior and alters the process model such that the selected process behavior is described by the resulting model (see \textbf{Extend} in \autoref{fig:overview}).
By incrementally selecting process behavior, the user \emph{guides the incremental process discovery algorithm} by providing feedback on the correctness of the observed event data.
The user therefore actively selects the process behavior to be added.
Since the incremental process discovery approach allows users to undo/redo steps at any time, they have more control over the process discovery phase of the model compared to conventional approaches. 
To improve the flexibility of Cortado, a process model editor is also embedded, allowing the user to alter the process model at any time (see \textbf{Edit} in \autoref{fig:overview}).
Furthermore, feedback mechanisms are implemented that notify the user of the quality of the discovered process models  (see \textbf{Compare} in \autoref{fig:overview}).

The remainder of this paper is structured as follows.
In \autoref{sec:background}, we briefly introduce background knowledge.
In \autoref{sec:algorithmic_foundation}, we explain the algorithmic foundation of Cortado, i.e., the incremental process discovery approach.
In \autoref{sec:functionality}, we present our tool and explain its main functionality and usage.
In \autoref{sec:implementation}, we briefly describe the underlying implementation.
\autoref{sec:conclusion} concludes the paper.

\section{Background}
\label{sec:background}
In this section, we briefly explain the concept of event data and present \emph{process trees}, which is the process modeling formalism used by Cortado.

\subsection{Event Data}
The information systems used in companies, e.g., Customer Relationship Management (CRM) and Enterprise Resource Planning (ERP) systems, track the performed activities during the executions of a process in great detail.

\autoref{tab:event_log} presents a simplified example of such event data, i.e., referred to as an \emph{event log}.
\begin{table}[tb]
    \rowcolors{1}{lightgray!30}{white}
  \centering
  \caption{Example (simplified) event data, originating from the \emph{Road Traffic Fine Management Process Event Log}~\cite{4TU:data/road}.
  Each row records an activity executed in the context of the process. The columns record various data related to the corresponding fine and the activity executed.}
  \label{tab:event_log}%
    \resizebox{\textwidth}{!}{
        \begin{tabular}{|c|c|c|c|c|c|c|c|c|c|c|}
        \hline
        \emph{\textbf{Fine}}  & \emph{\textbf{Event}} & \emph{\textbf{Start}} & \emph{\textbf{Complete}} & \emph{\textbf{Amount}} & \emph{\textbf{Notification}} & \emph{\textbf{Expense}} & {\emph{\textbf{Payment}}} & {\emph{\textbf{Article}}} & \makecell{\emph{\textbf{Vehicle}}\\ \emph{\textbf{ Class}}} & \makecell{\emph{\textbf{Total}}\\\emph{\textbf{Payment}}} \\\hline\hline
        A1    & Create Fine & 2006/07/24  & 2006/07/24  & 35.0  &       &       &          & 157          & A     & 0.0 \\
        A1    & Send Fine & 2006/12/05  & 2006/12/05  & 35.0  &       & 11.0  &             & 157          & A     & 0.0 \\
        A100  & Create Fine & 2006/08/02  & 2006/08/02  & 35.0  &       &       &          & 157          & A     & 0.0 \\
        A100  & Send Fine & 2006/12/12  & 2006/12/12  & 35.0  &       & 11.0  &            & 157          & A     & 0.0 \\
        A100  & \makecell{Insert Fine\\Notification} & 2007/01/15  & 2007/01/15  & 35.0  & P     & 11.0      &       & 157          & A     & 0.0 \\
        A100  & Add penalty & 2007/03/16  & 2007/03/16  & 71.5  & P     & 11.0      &       & 157          & A     & 0.0 \\
        A100  & \makecell{Send for Credit\\Collection} & 2009/03/30  & 2009/03/30  & 71.5  & P     & 11.0      &       & 157          & A     & 0.0 \\
        A10000 & Create Fine & 2007/03/09  & 2007/03/09  & 36.0  &       &       &         & 157          & A     & 0.0 \\
        A10000 & Send Fine & 2007/07/17  & 2007/07/17  & 36.0  &       & 13.0  &             & 157          & A     & 0.0 \\
        A10000 & Add penalty & 2007/10/01  & 2007/10/01  & 74.0  & P     & 13.0      &       & 157          & A     & 0.0 \\
        A10000 & Payment & 2008/09/09  & 2008/09/09  & 74.0  & P     & 13.0       & 87.0 & 157          & A     & 87.0 \\
        $\dots$&$\dots$&$\dots$&$\dots$&$\dots$&$\dots$&$\dots$&$\dots$&$\dots$&$\dots$&$\dots$ \\ \hline
        \end{tabular}%
        }
\end{table}%
Each row represents an \emph{event}, a recording related to some \emph{activity instance} of the process.
For example, the first row indicates that a fine with identifier \texttt{A1} was created on July 24, 2006.
The next line/event records that the same fine was sent.
Note that the corresponding expense for sending the fine was \euro{}11.0, the Article of this violation is 157, the vehicle class is A, etc.
Multiple rows have the same value for the \emph{Fine}-column, i.e., often referred to as the \emph{case identifier}; all these events are executed for the same instance of the process, e.g., for the same customer, the same patient, the same insurance claim, or, in the given case, for the same fine.
We refer to the digital recording of a process instance as a \emph{case}.
As such, an \emph{event log}, describes a \emph{collection of cases}.
In Cortado, we focus on \emph{trace variants}, i.e., unique sequences of executed activities.
For instance, for the fine A1 we observe the trace $\langle$\emph{Create Fine, Send Fine}$\rangle$ and for the fine A100 $\langle$\emph{Create Fine, Send Fine, Insert Fine Notification, Add penalty, Send for Credit Collection}$\rangle$.
Note that, in general, there may be several cases for which the same sequence of activities has been performed.

\subsection{Process Trees}
We use \emph{process models} to describe the control-flow execution of a process.
Some \emph{process modeling formalisms} additionally allow specifying, for instance, what resources execute an activity and what data attributes in the information system might be read or written during the activity execution.
In Cortado, we use \emph{process trees} as a process modeling formalism.
Process trees are a hierarchical process modeling notation that can be expressed as \emph{sound Workflow nets} (sound WF-nets), i.e., a subclass of Petri nets, often used to model business processes.
Process trees are annotated rooted trees and correspond to the class of \emph{block-structured WF-nets}, a subclass of sound WF-nets.
Process trees are used in various process discovery algorithms, e.g., the inductive miner~\cite{DBLP:conf/apn/LeemansFA13}.

\begin{figure}[tb]
     \centering
     \begin{subfigure}[b]{1\textwidth}
        \centering
        \includegraphics[width=\textwidth]{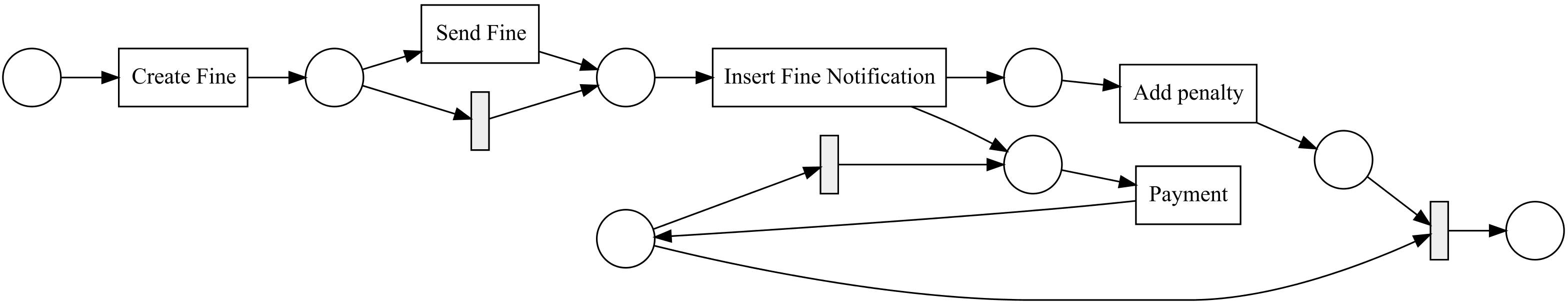}
        \caption{Simple example Petri net (sound WF-net) modeling a road fine management process.}
        \label{fig:petri_net}
     \end{subfigure}~
     
     \begin{subfigure}[b]{1\textwidth}
        \centering
        \includegraphics[width=\textwidth]{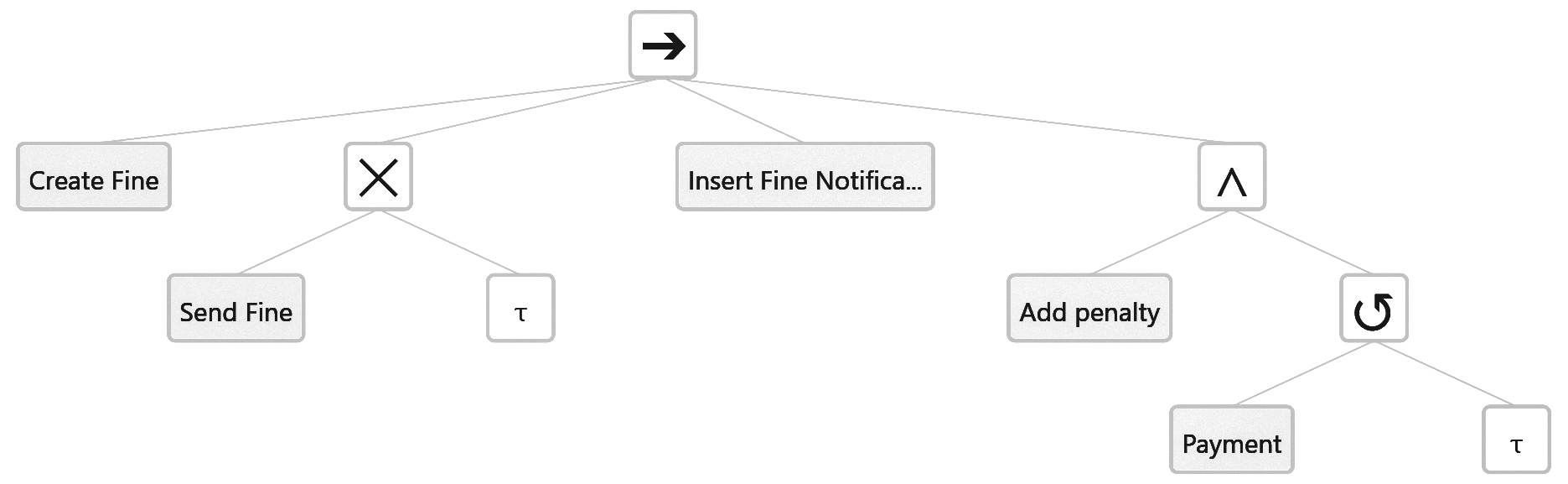}
        \caption{A process tree modeling the same behavior as the Petri net in \autoref{fig:petri_net}.}
        \label{fig:process_tree}
     \end{subfigure}~
     
     \caption{Two process models, a Petri net (\autoref{fig:petri_net}) and a process tree (\autoref{fig:process_tree}), describing the same process behavior, i.e., a simplified fine management process}
     \label{fig:models}
\end{figure}

In \autoref{fig:models}, we show two simplified models of the road fine management process, which is partially shown in \autoref{tab:event_log}.
\autoref{fig:petri_net} shows a sound WF-net.
\autoref{fig:process_tree} shows a process tree describing the same behavior as the model in \autoref{fig:petri_net}.
Both models describe that the \emph{Create Fine} activity is executed first.
Secondly, the \emph{Send Fine} activity is optionally executed.
Then, the \emph{Insert Fine Notification} activity is performed, followed by a block of concurrent behavior including \emph{Add penalty} and potentially multiple executions of \emph{Payment}.

The semantics of process trees are fairly simple, and, arguably, their hierarchical nature allows one to intuitively reason about the general process behavior.
Reconsider \autoref{fig:process_tree}.
We refer to the internal vertices as \emph{operators} and use them to specify control-flow relations among their children.
The leaves of the tree refer to \emph{activities}.
The \emph{unobservable activity} is denoted by $\tau$.
In terms of operators, we distinguish four different types: the \emph{sequence operator ($\to$)}, the \emph{exclusive choice operator ($\times$)}, the \emph{parallel operator ($\wedge$)}, and the \emph{loop operator ($\circlearrowleft$)}.
The sequence operator ($\to$) specifies the execution of its subtrees in the given order from left to right.
The exclusive choice operator $(\times)$ specifies that \emph{exactly one} of its subtrees gets executed.
The parallel operator $(\wedge)$ specifies that all subtrees get executed in any order and possibly interleaved.
The loop operator $(\circlearrowleft)$ has exactly two subtrees.
The first subtree is called the \enquote{do-part}, which has to be executed at least once. 
The second subtree is called the \enquote{redo-part}, which is optionally executed. 
If the redo-part gets executed, the do-part is required to be executed again.

\section{Algorithmic Foundation}
\label{sec:algorithmic_foundation}
\begin{figure}[tb]
    \centering
    \includegraphics[width=.95\textwidth, clip, trim=0 5cm 5cm 0 ]{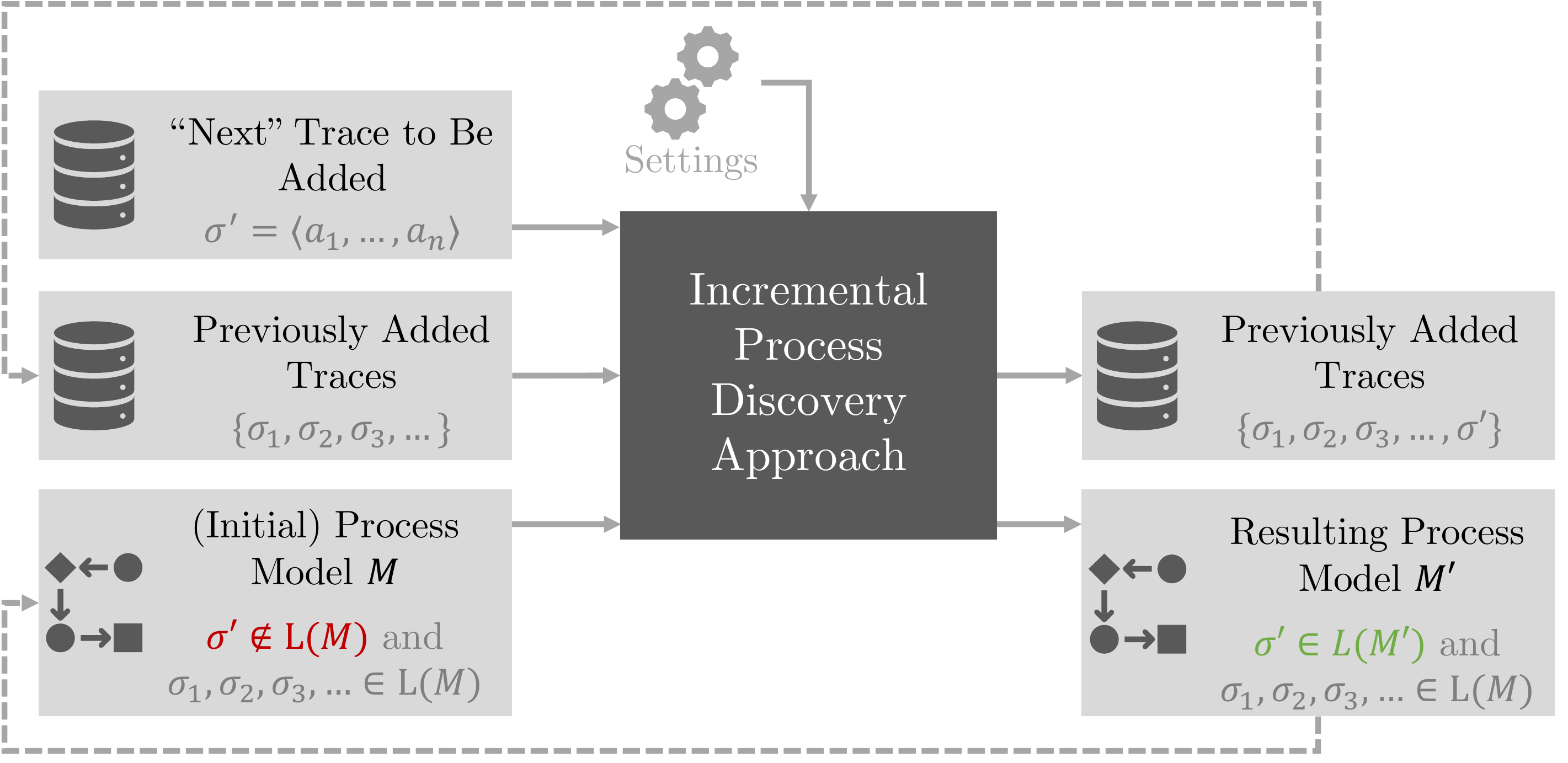}
    \caption{Schematic overview of incremental process discovery (presented in our earlier work~\cite{10.1007/978-3-030-50316-1_25}), i.e., the algorithmic foundation of Cortado. Starting with an initial process model $M$ and observed process behavior (a trace $\sigma'$ capturing a sequence of executed process activities: $a_1,\dots,a_n$) that is not yet captured by the model, the incremental discovery approach alters the given process model $M$ into a new model $M'$ that additionally accepts the given trace $\sigma'$}
    \label{fig:overview_incr_discovery}
\end{figure}

In this section, we briefly describe the algorithmic foundation of Cortado's incremental process discovery approach~\cite{10.1007/978-3-030-50316-1_25}.
Consider \autoref{fig:overview_incr_discovery}, in which we present a schematic overview on said algorithmic foundation.

As an input, we assume a process model $M$, which is either given initially or the result of a previous iteration of the incremental discovery algorithm.
Additionally, a \emph{trace} $\sigma'{=}\langle a_1,\dots,a_n \rangle$, i.e., a sequence of executed activities $a_1,\dots,a_n$, is given.
We assume that the trace $\sigma'$ is not yet part of the language of model $M$ (visualized as $\sigma'{\notin}L(M)$).
Note that $\sigma'$ is selected by the user.
If the incremental procedure has already been executed before, i.e., traces have been already added to the process model in previous iterations, we use those traces as input as well (visualized as $\{\sigma_1,\sigma_2,\sigma_3,\dots\}$ in \autoref{fig:overview_incr_discovery}).
The incremental process discovery algorithm transforms the three input artifacts into a new process model $M'$ that describes the input trace $\sigma'$ and the previously added traces $\{\sigma_1,\sigma_2,\sigma_3,\dots\}$.
In the next iteration, the user selects a new trace $\sigma''$ to be added and the set of previously added traces gets extended, i.e., $\{\sigma_1,\sigma_2,\sigma_3,\dots\}{\cup}\{\sigma'\}$.

As mentioned before, Cortado uses process trees as a process model formalism. 
The incremental discovery approach~\cite{10.1007/978-3-030-50316-1_25} exploits the hierarchical structure of the input process tree $M$ and pinpoints the subtrees where the given trace $\sigma'$ deviates from the language described from the model.
To identify the subtrees, the process tree is converted into a Petri net and alignments~\cite{https://doi.org/10.1002/widm.1045} are calculated.
Subsequently, the identified subtrees get locally replaced, i.e., $M'$ is a locally modified version of $M$.

\section{Functionalities and User Interface}
\label{sec:functionality}
\begin{figure}[tb]
    \centering
    \includegraphics[width=\textwidth]{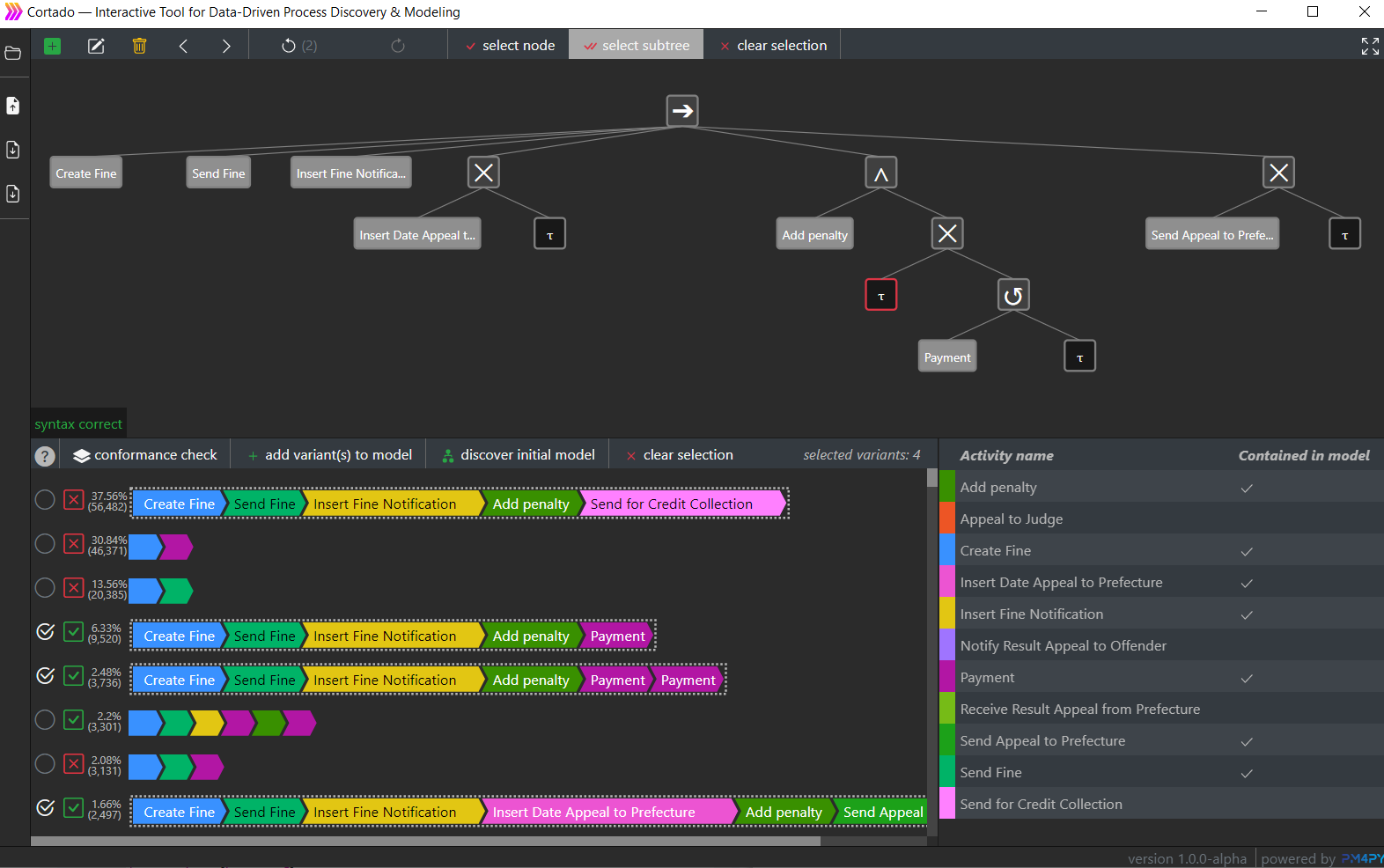}
    \caption{Screenshot of the graphical user interface of Cortado. In the screenshot, we have loaded the \emph{Road Traffic Fine Management Process Event Log}~\cite{4TU:data/road}.}
    \label{fig:ui}
\end{figure}

In this section, we present the main functionalities of Cortado.
We do so along the lines of the user interface of Cortado as visualized in \autoref{fig:ui}.

\subsection{I/O Functionalities}
Cortado supports various importing and exporting functionalities, which can be triggered by the user by clicking the import/export buttons visualized in the left sidebar, see \autoref{fig:ui}.
Cortado supports importing event data stored in the IEEE eXtensible Event Stream (XES) format~\cite{XES}.
Furthermore, Cortado supports importing process tree models stored as a \texttt{.ptml}-file, for instance, if an initial (manual) process model is available.
Process tree model files (\texttt{.ptml}-files) can be generated, e.g., by process mining tools such as ProM\footnote{\url{https://www.promtools.org}} and PM4Py\footnote{\url{https://pm4py.fit.fraunhofer.de/}}.

Next to importing, Cortado supports exporting the discovered process model both as a Petri net (\texttt{.pnml}-file) and as a process tree (\texttt{.ptml}-file).
In short, Cortado offers a variety of I/0 functionalities and, hence, can be easily combined with other process mining tools.

\subsection{Visualizing and Editing Process Trees}

\begin{figure}[tb]
    \centering
    \includegraphics[width=\textwidth]{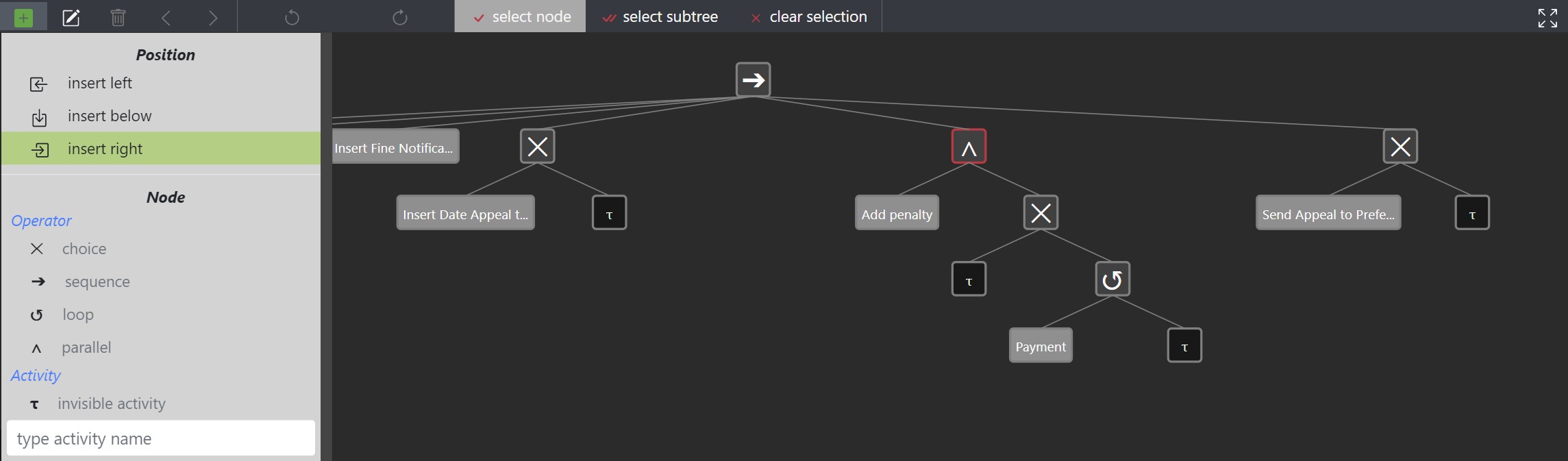}
    \caption{Screenshot of the process tree editor in Cortado}
    \label{fig:tree-editor}
\end{figure}

Cortado supports the visualization and editing of process trees.
The \enquote{process tree under construction} -- either loaded or iteratively discovered -- is visualized in the upper half of the tool (\autoref{fig:ui}).
The user can interactively select subtrees or individual vertices of the process tree by clicking an operator or a leaf node.
Various edit options, e.g., removing or shifting the selected subtree left or right, are available from the top bar of the application (\autoref{fig:ui}).
Apart from removing and shifting subtrees, the user can also add new subtrees to the process tree.
\autoref{fig:tree-editor} shows a screenshot of the tree editor in detail.
In the given screenshot, an inner node, a parallel operator ($\wedge$), is selected.
Based on the selected inner node, the user can specify the position where to add a new node in the dropdown-menu by clicking on either \texttt{insert left}, \texttt{insert right} or \texttt{insert below}.
In the given screenshot, \texttt{insert right} is selected.
Next, the user can choose between an activity (a leaf node) or an operator (an inner node).
By clicking on one of the options, the new node is added directly to the right of the selected node.
In summary, the process tree editor in Cortado allows the user to alter the process tree at any time.

\subsection{Event Data Interaction}

\begin{figure}[tb]
    \centering
    \includegraphics[width=.95\textwidth]{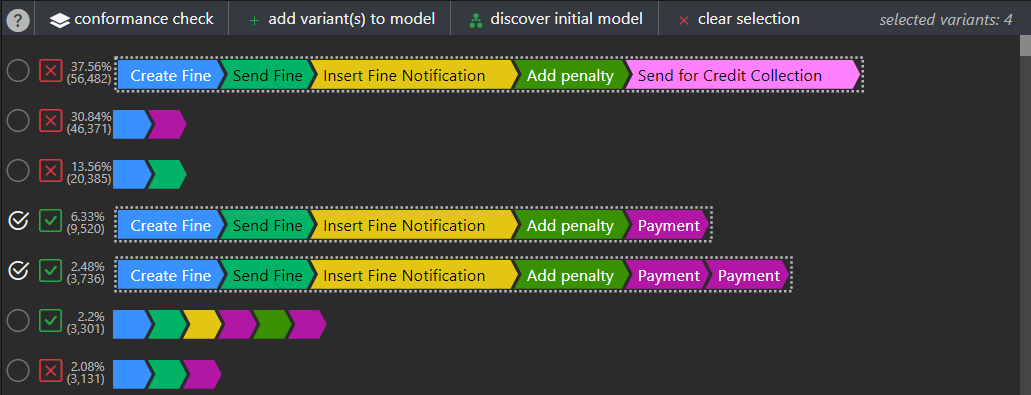}
    \caption{Screenshot of the trace variants visualization in Cortado. There are two icons to the left of each trace variant. The left icon, a circle or a circle with a check mark, indicates whether a trace variant has been explicitly added to the model by the user. The right icon, a red cross or a green check mark, indicates if the trace variant is accepted by the current process model}
    \label{fig:variant-explorer}
\end{figure}

To visualize the loaded event data, Cortado uses trace variants.
Clearly, multiple instances of a process can describe the exact same behavior regarding the sequence of observed activities.
For example, the most frequently observed sequence of behavior in the \emph{Road Traffic Fine Management Process Event Log~\cite{4TU:data/road}} (i.e., used in \autoref{fig:ui}), describes the sequence: $\langle$\emph{Create Fine}, \emph{Send Fine}, \emph{Insert Fine Notification}, \emph{Add Penalty}, \emph{Send for Credit Collection}$\rangle$.
In total, the process behavior of 56,482 fines (37.56\% of the total number of recorded fines) follows this sequence of activities.

Trace variants are visualized in Cortado as a sequence of \emph{colored chevrons}.
Each activity gets a unique color assigned.
For instance, the activity \emph{Create Fine} is assigned a blue color in \autoref{fig:variant-explorer}.
Cortado sorts the trace variants based on their frequency of occurrence, descending from top to bottom.
By clicking a trace variant, the user \enquote{selects a variant}.
Selection of multiple variants is also supported.
In case an initial model does not exist, clicking the \texttt{discover initial model} button discovers one from the selected trace variants using the Inductive Miner~\cite{DBLP:conf/apn/LeemansFA13}, a process discovery algorithm that guarantees replay fitness on the given traces and returns a process tree.
In case an initial model is present, the selected variants can be \enquote{added to the model} by clicking the \texttt{add variant(s) to model} button. 
In this case, Cortado performs incremental process discovery as described in~\autoref{sec:algorithmic_foundation}.

Left to each trace variant, we see statistics about its occurrence in the event log and two icons.
The left-most icon, an empty circle or a white check mark, indicates whether or not the trace variant has been explicitly added to the model by the user (\autoref{fig:variant-explorer}).
A variant has been explicitly added by the user if either the variant was used to discover an initial model or the variant has been added to an existing model by applying incremental discovery, i.e., the variant was selected and the user pressed the button \texttt{add variant(s) to model}.
Note that it is possible that a particular trace variant which was not explicitly selected by the user is described by the process model; however, after incrementally adding further variants to the model, the variant is potentially no longer described.
In contrast, Cortado guarantees that explicitly added trace variants are always described by any future model incrementally discovered.
However, since Cortado allows for manual tree manipulation at any time, it might be the case that an explicitly added variant is not described anymore by the tree due to manual changes to the process tree.

The right-most icon is either a red cross or a green check mark (\autoref{fig:variant-explorer}).
These icons indicate whether a trace variant is described/accepted by the process model, i.e., if a trace variant is in the language of the process model.
For the computation of these conformance statistics, we use \emph{alignments}~\cite{https://doi.org/10.1002/widm.1045}.
Therefore, we internally translate the process tree into a Petri net and execute the alignment calculation.
For instance, the first three variants in \autoref{fig:variant-explorer} are not accepted by the current process model, but the last two variants are accepted.
Similar to triggering incremental discovery, manipulations of the process tree potentially result in variants that are no longer described by the process model.
To assess the conformity of traces after editing the process tree manually, the user can trigger a new conformity check by clicking the \texttt{conformance check} button.

Lastly, Cortado shows an overview list of all activities from the loaded event log. 
This overview is located in the lower right part of Cortado's user interface (\autoref{fig:ui}).
Besides listing the activity names, Cortado indicates -- by using a check mark icon -- which activities from the event log are already present in the process model under construction.
Thereby, the user gets a quick overview of the status of the incremental discovery.


\section{Implementation and Installation}
\label{sec:implementation}

The algorithmic core of Cortado is implemented in Python.
For the core process mining functionality, we use the PM4Py\footnote{\url{https://pm4py.fit.fraunhofer.de/}} library, a python library that contains, for instance, event log handling and conformance checking functionality.
The GUI is implemented using web technologies, e.g., we chose the Electron\footnote{\url{https://www.electronjs.org/}} and Angular\footnote{\url{https://angular.io/}} framework to realize a cross-platform desktop application.
For the graphical representation of the process tree and the trace variants we use the JavaScript library d3.js\footnote{\url{https://d3js.org/}}.

The tool is available as a desktop application and can be freely downloaded at \url{https://cortado.fit.fraunhofer.de/}.
The provided archive, a \texttt{ZIP}-file, contains an executable file that will start the tool.
Upon starting, the data used within this paper, i.e., Road Traffic Fine Management Process~\cite{4TU:data/road}, gets automatically loaded. 
Moreover, the archive contains examples of other event logs available as \texttt{XES}-files in the directory \texttt{example\_event\_logs}.


\section{Conclusion and Future Work}
\label{sec:conclusion}

This paper presented Cortado, a novel tool for interactive process discovery and modeling.
The tool enables the user to incrementally discover a process model based on observed process behavior.
Therefore, Cortado allows to load an event log and visualizes the trace variants in an intuitive manner. 
Starting from an initial model, which can be either imported or discovered, the user can incrementally add observed process behavior to the process model under construction.
Various feedback functionalities, e.g., conformance checking statistics and the activity overview, give the user an overview of the process model under construction anytime.
Supporting common file formats such as XES and PNML, Cortado can be easily used with other process mining tools.

In future work, we plan to extend Cortado's functionality in various ways. 
First, we aim to offer more options for the user to interact with the underlying incremental discovery approach. 
For example, we plan to allow the user to \emph{lock} specific subtrees during incremental discovery to prevent these from being modified further.
We also plan, in case the user changes the tree in the editor, to provide improved and instant feedback on the conformance impact the changes have w.r.t. the loaded event log and the already explicitly added trace variants. 
However, since the calculation of conformance checking statistics -- a crucial part for instant user feedback -- is computationally complex, we plan to evaluate the extent to which approximation algorithms~\cite{schuster2020alignment} can be integrated. 

Next to further functionality, we plan to conduct case studies with industry partners.
Thereby, we aim to focus on the practical usability of Cortado.
The goal is to investigate which interaction options are meaningful and understandable for the user interacting with Cortado.

%
%
%
\bibliographystyle{splncs04}
\bibliography{main}

\end{document}